\documentclass[runningheads]{llncs}
\usepackage{graphicx}
\usepackage{comment}
\usepackage{float}
\usepackage{subcaption}
\usepackage{hyperref}
\begin{document}
\title{Behavior of Keyword Spotting Networks Under Noisy Conditions
\thanks{This work has been partly funded by the European Union (EU) and the German Federal Ministry of Education and Research (BMBF) in the project OCEAN12 (reference number: 16ESE0270).\\
The final authenticated publication is available online at \url{https://doi.org/10.1007/978-3-030-86362-3_30}}
}
%
%
\author{Anwesh Mohanty\inst{1} \orcidID{0000-0001-9257-5175} \and
Adrian Frischknecht\inst{2} \orcidID{0000-0002-1795-0948} \and
Christoph Gerum\inst{2} \orcidID{0000-0002-1715-567X} \and
Oliver Bringmann\inst{2} \orcidID{0000-0002-1615-507X}}
\authorrunning{A. Mohanty et al.}
%
\institute{Indian Institute Technology Bombay, Mumbai 400076, India \and
University of Tübingen, 72074 Tübingen, Germany
}
\maketitle              
\begin{abstract}
Keyword spotting (KWS) is becoming a ubiquitous need with the advancement in artificial intelligence and smart devices. Recent work in this field have focused on several different architectures to achieve good results on datasets with low to moderate noise. However, the performance of these models deteriorates under high noise conditions as shown by our experiments. In our paper, we present an extensive comparison between state-of-the-art KWS networks under various noisy conditions. We also suggest adaptive batch normalization as a technique to improve the performance of the networks when the noise files are unknown during the training phase. The results of such high noise characterization enable future work in developing models that perform better in the aforementioned conditions. 

\keywords{Keyword spotting  \and High noise conditions \and Adaptive batch normalization \and Sinc convolution network \and Temporal convolution ResNet.}
\end{abstract}
\section{Introduction}

Automatic speech recognition is one of the fastest developing fields in artificial intelligence and machine learning. With the advent of smart assistants (e.g. Google assistant, Siri, Cortana) in most of the latest devices, the ability of speech recognition software to recognize certain wake words (e.g. ‘‘Ok Google’’, ‘‘Hey Siri’’) from continuous speech filled with varying levels of background noise becomes paramount in enhancing the user experience. 

The networks used for KWS have evolved significantly from the initial Gaussian Mixture Model-Universal Background Models (GMM-UBMs) and Hidden Markov Models (HMMs) to Deep Neural Networks (DNNs) to the current use of different variations of Convolutional Neural Networks (CNNs) ~\cite{abdel2014convolutional,chen2014small,hinton2012deep,sainath2015convolutional}. Due to their inherent properties, CNNs can discover robust and invariant representations of the input waveforms provided to them, and have obtained state-of-the-art performance on several speech recognition tasks carried out under moderate noise conditions.

\begin{figure}[H]
\centering
\includegraphics[scale=0.3]{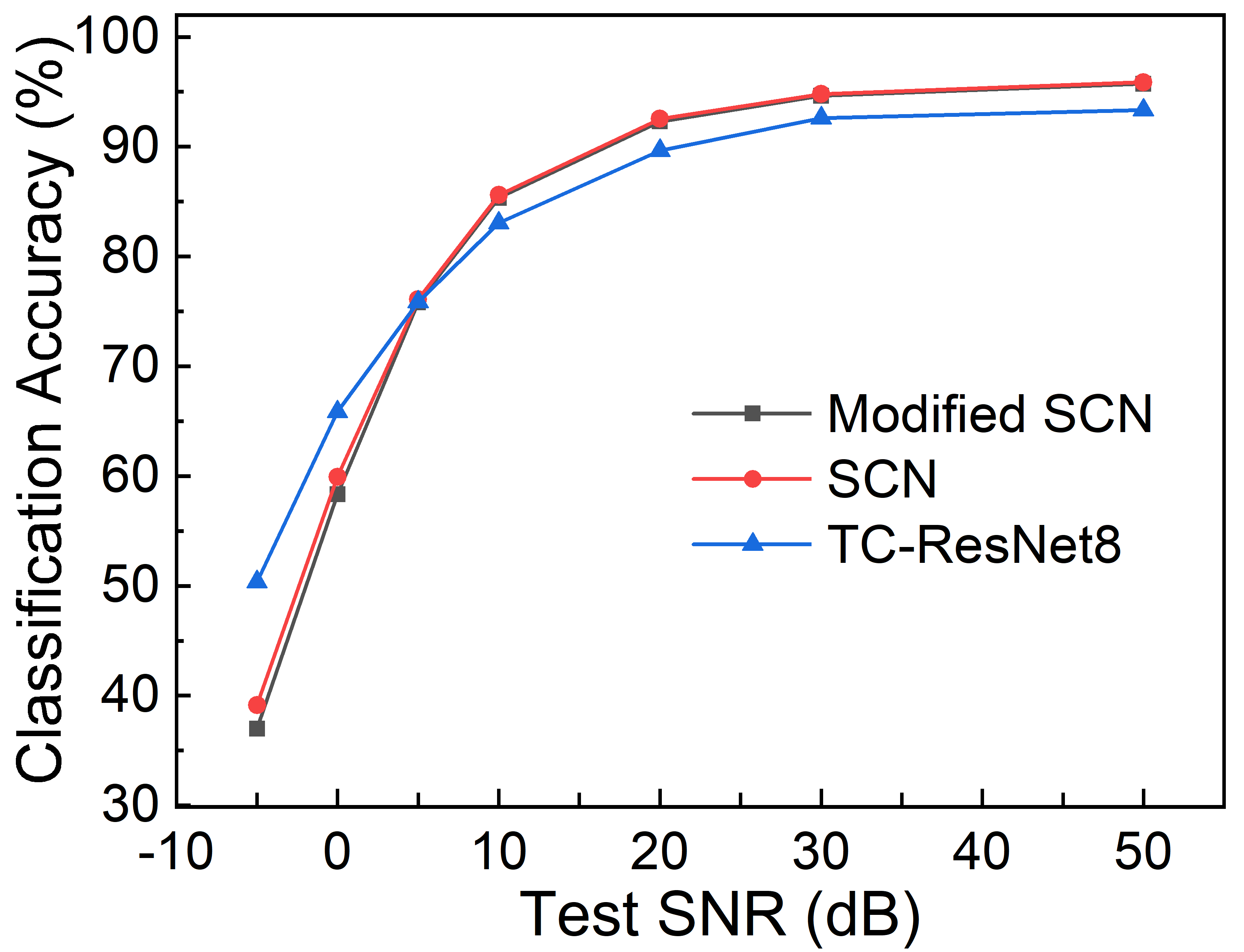}
\caption{Performance of networks under varying noise (trained on clean dataset). Classification accuracy of all networks decreases significantly as the level of noise increases beyond a certain threshold.}
\label{fig_sim}
\end{figure}

KWS networks have undergone several transformations in their architectures and currently CNNs provide the best performance under moderate noise conditions. Despite their impressive performance, as demonstrated in Fig. \ref{fig_sim}, the test accuracy falls at a steep rate once the signal-to-noise ratio (SNR) in the dataset crosses a certain threshold. Considering day-to-day situations like heavy traffic, construction sites etc, the places where there is very high background noise, the current architectures won't give the same performance as they will give in a lower noise environment. This calls for an architecture which can perform the task of KWS under such noisy conditions with a competitive accuracy.

For our experiments, we use three models for comparison. The first one is the TC-ResNet8 (TC: temporal convolutions) which uses pre-processed MFCC features as inputs for classification. The second one is the SincConv Network (SCN) which classifies on raw audio data and finally the last model is our variation of the SCN but with optimal parameter tuning to reduce the memory footprint and total computation cost without reducing the classification accuracy. We subject the models to different noise conditions to obtain a detailed characterization of the models' performances. 

The remainder of the paper is organized as follows. Sec.2 gives a brief description of other relevant work going on in this field. Sec.3 discusses the basic features of all the architectures used during evaluation. Sec.4 outlines the experimental setup and the results respectively. Here we also propose the use of batch normalization (BatchNorm) as a method to adapt the network to unknown noise conditions. Finally, Sec.5 discusses our conclusions and scope for future work. 

\section{Related Works}

Significant research has been done in the field of KWS in recent times, with a focus on developing compact and accurate models that can be implemented in hardware without consuming too much power. Zhang et al. \cite{Zhang:17} provides a comparison of performance and hardware requirement (memory and operation count) of Deep Neural Network (DNN), CNN, Long short-term memory (LSTM), and depthwise separable (DS) CNN models on MFCC feature data as input, where the DS-CNN provides the best result. Choi et al. \cite{Choi} proposed the TC-ResNet which provides state-of-the-art 96.6\% accuracy on MFCC input data, as well as a speedup of 385$\times$ compared to previous architectures on the Google Speech Commands Dataset~\cite{dataset}. Since pre-processed data like MFCC features won't be always available, few CNN architectures have been developed to work on raw audio data as input. One of the notable ones is the SCN architecture proposed by Mittermaier et al.~\cite{Mittermaier:19}, which uses SincNet \cite{Ravanelli} and DS convolutions~\cite{dsconv} to achieve comparable accuracy to the state-of-the-art TC-ResNet models. 

There is very little documentation about the performance of popular KWS networks under high noise, or in situations where the noise present during the inference stage is much different from that during training. Liu et al. \cite{Liu} have provided a brief noise characterization of the performance of their binary weight network using different types of noise like white, pink and miscellaneous noise in daily-life activities. Huang et al. \cite{Huang}, Raju et al. \cite{Raju} and Pervaiz et al. \cite{Pervaiz} have provided detailed studies on the performance of their systems for the task of KWS under noise, but the datasets and the metrics used in these works are all different and cannot be used to draw a comparison with current state-of-the-art models. To the best of our knowledge, we are the first to provide a detailed characterization of popular KWS networks on a standard dataset under varying high noise conditions and provide a simple and efficient solution to improve the accuracy by quite a significant margin in the aforementioned conditions. 

\section{Model Architectures}

\begin{figure}[H]
\centerline{\includegraphics[width = \textwidth]{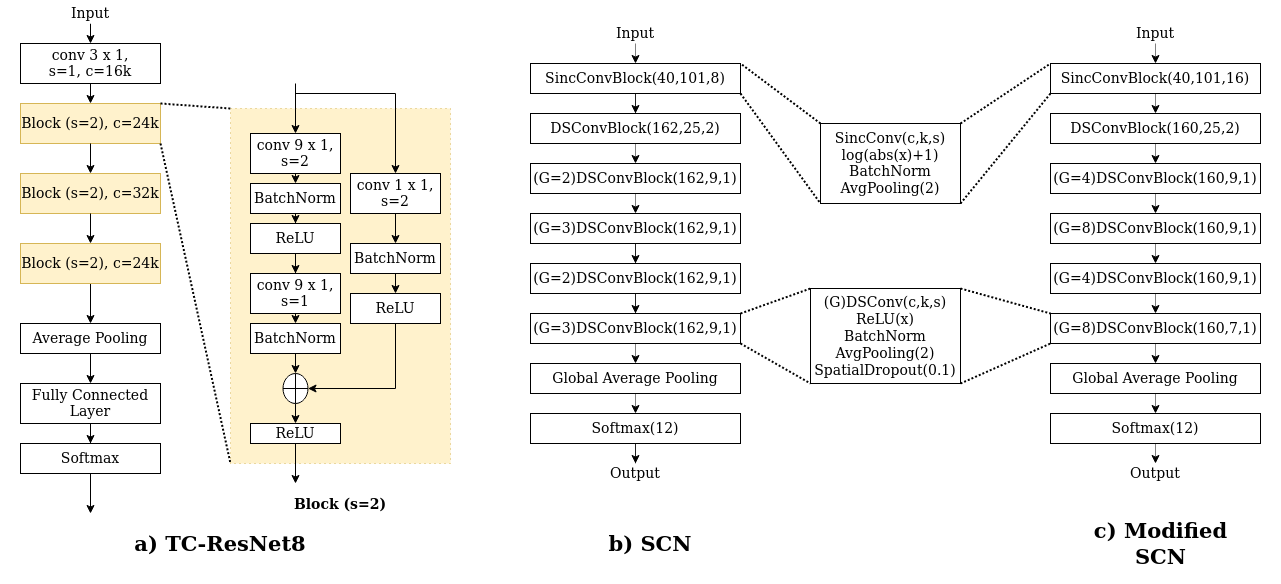}}
\caption{Architectures of the models. The hyperparameters \textit{c, k} and \textit{s} represent the number of output channels, kernel size and stride respectively for all the models. Architectures of TC-ResNet8 and SCN adopted from~\cite{Choi} and~\cite{Mittermaier:19} respectively.}
\label{fig:model_arch}
\end{figure}

In this paper we consider three representative neural networks. Table 1 summarizes the multiply accumulate operations (MACs) and total weights in the considered models and Fig. \ref{fig:model_arch} shows the respective architectures.
\begin{table}[H]
\centering
\caption{\label{font-table} Summary of Models }
\begin{tabular}{|c|c|c|}
\hline \textbf{Model} &  \hspace{0.1cm} \textbf{MACs} \hspace{0.1cm} & \hspace{0.1cm}\textbf{Parameters} \hspace{0.1cm} \\ \hline
TC-ResNet8  & 1.5M & 66k\\
SCN & 18M & 60k \\
\hspace{0.1cm} Modified SCN \hspace{0.1cm} & 7.5M & 34.5k\\
\hline
\end{tabular}

\end{table}

\subsection{TC-ResNet8 Architecture}

TC-ResNet8 (Fig. \ref{fig:model_arch})~\cite{Choi} is a CNN architecture which utilizes temporal convolutions, 1-D convolutions along the temporal dimension, for KWS and classifies on the MFCC data (pre-processed from raw audio signals) as input. This model adopts ResNet~\cite{Resnet}, one of the most popular CNN architectures, but uses $m \times 1$ kernels (m=3 for the first layer and m=9 for the other layers) in its layers. By switching to temporal convolutions instead of 2D convolutions, there is a decrease in the output feature map of each layer which leads to a huge reduction in the computational burden and memory footprint of the subsequent layers. 

TC-ResNet8 model has shown good performance for KWS with only 66k parameters. There is no bias in the convolution and fully connected layers. Each batch normalization layer has trainable parameters for scaling and shifting. The TC-ResNet8 model has 3 residual blocks and \{16, 24, 36, 48\} channels for each layer including the first convolution layer.

\subsection{SCN Architecture}

SCN network (Fig. \ref{fig:model_arch}) \cite{Mittermaier:19} uses rectangular band-pass filters (in the frequency domains) in the first convolutional layer to classify on the input raw audio waveform. This is equivalent to convolving the input signal with parametrized sinc functions ($\mbox{sinc}(x)= \frac{\sin(x)}{x}$) in the time domain. The filters can be represented as:
\begin{equation}
    H[f,f_1,f_2]= \mbox{rect}(\frac{f}{f_2})-\mbox{rect}(\frac{f}{f_1})\label{eq}
\end{equation}
\begin{equation}
    h[n,f_1,f_2]= 2f_{2}\mbox{sinc}(2\pi f_{2}n)-2f_{1}\mbox{sinc}(2\pi f_{1}n)\label{eq}
\end{equation}

From (1), the frequency domain expression of the filters, we can see that a single filter extracts only the information lying between the two frequency levels, $f_{1}$ and $f_{2}$. This extracted data acts as a feature set for the consequent CNN layers. Since only two parameters, the upper and lower cut-off frequencies, are required to define any sinc filter, this leads to a smaller memory footprint. As suggested in \cite{Mittermaier:19}, a log-compression activation ($y=\log(\mbox{abs}(x)+1)$) is used after the sinc convolutions.

In the subsequent layers we have five grouped DS convolutional blocks. DS convolutions \cite{dsconv} are a great alternative to standard convolutions as they reduce the computation power by a significant value without reducing the effectiveness much. Grouping \cite{group} is introduced to reduce the number of parameters introduced by the pointwise convolutions after each depthwise convolution. Each convolution block is followed by layers for batch normalization, spatial dropout for regularization and an average pooling block. After these 5 blocks we have a global average pooling block followed by a softmax layer to obtain the class posteriors to classify into 12 classes. 

\subsection{Modified SCN Architecture}

As can be seen from Table \ref{font-table}, though there is a decrease in the parameter count when we go from the TC-ResNet8 to the SCN model as well as the added benefit of not spending resources on pre-processing to obtain the MFCC data, the number of MACs inside the latter model is almost 12 times more than the former model (excluding the MACs in pre-processing of raw  audio in TC-ResNet8). This huge level of disparity in the computation costs of the two models certainly raises questions over the viability of the SCN network over the TC-ResNet8 while considering a hardware implementation.

Following several experiments, study of the properties of the SCN model and extensive fine-tuning of the hyperparameters, we present the modified SCN model (Fig. \ref{fig:model_arch}) which gives comparable accuracy to the original model but reduces the computation cost and memory footprint by almost a factor of two. The subtle changes can be seen in Fig. \ref{fig:model_arch} and Table 1. We change the grouping in CNN layers from alternate (2,3) grouping to alternate (4,8) grouping. This impacts primarily the total number of parameters used in the model as can be seen in Table 1. Almost 50\% of the computations are carried out in the very first sinc convolution layer. To tackle this issue we double the stride in the sinc convolution layers which leads to decrease in the MACs in the subsequent layers as well. 

The primary motivation of the modified SCN architecture is to obtain the best possible optimized version of the SCN architecture without sacrificing any of the advantage the SCN network has over the TC-ResNet8 architecture. The difference in performance due to the architecture changes in Fig. \ref{fig:model_arch} won't be noticeable when running the networks on a modern GPU, but on moving the networks to smaller embedded systems for a more practical setup, there will be a significant change in the latency and power consumption due to the discrepancy in the number of MACs and parameters compared to the SCN network. Based on the optimizations that we have incorporated into the SCN architecture, the modified SCN is able to compete with both the SCN and TC-ResNet8 networks in terms of accuracy and efficiency respectively.

\subsection{Batch Normalization Method}

In standard neural networks problems, the statistics of batches during training are learned in the BatchNorm layers and used without changing during the test and validation phases. This works well in most cases because the test statistics resemble closely to the training statistics. But when the test statistics vary significantly from the training statistics due to environmental noise, this assumption fails and the model won't perform well. In this case, a better training dataset should be found, but that is not always possible. To tackle this fall in performance and not having to resort to finding a new training set, we adapt a simple modification from Schneider et al.~\cite{Schneider:20} - we do not switch off the BatchNorm layers during the validation and test phases when the noise during test is unknown. This way the network will learn and use the batch norm statistics during inference rather than the training statistics which might vary significantly from the inference statistics, and provide us with better results as seen in Fig. \ref{fig:acc_batchnorm} without a significant computation overhead.

\section{Experimental Evaluation}

The networks mentioned in Sec.3 have been trained and evaluated on the Google’s Speech Commands dataset~\cite{dataset}. The dataset consists of 105,829 one-second (or less) long utterances of 35 different keywords spoken by 2,618 different speakers. We choose the following 10 keywords: ‘‘yes’’, ‘‘no’’, ‘‘up’’, ‘‘down’’, ‘‘left’’, ‘‘right’’, ‘‘on’’, ‘‘off’’, ‘‘stop’’, ‘‘go’’, along with classes for unknown and silence.  The remaining 25 words are labeled as unknown. The utterances are then randomly divided into training, validation and test sets in the ratio of 80:10:10 respectively.  

For noise injection, one-second chunks are chosen randomly from three types of noise present: white, pink and miscellaneous (consisting of samples from real life activities like traffic noises, conversation, flowing water etc.). For the training phase, these chunks are sampled randomly between the SNR range of [-5dB, +10dB] and added to the clean dataset. For the validation and test phase, the SNR value is kept fixed at one of the following values: -5dB, 0dB, +5dB, +10dB. To ensure that the final signal after mixing the noise with the dataset does not get clipped at any instant, we introduce a small gain block to scale the signals so that the SNR remains constant and no clipping takes place.


For the first experiment, all the noise files are available for the training, validation and test phases. For the second experiment, white and pink noise is injected into the training dataset and miscellaneous noise is added to validation and test dataset. Our model is trained for 150 epochs with the Stochastic Gradient Descent (SGD) optimizer with an initial learning rate of 0.1 and learning rate decay of 0.75 after 10 epochs. The model with highest validation accuracy after 150 epochs is saved to evaluate the accuracy on the test set.

\subsection{Network performance when noise conditions are known}

\begin{figure}[!htbp]
\centering
  \includegraphics[scale=0.3]{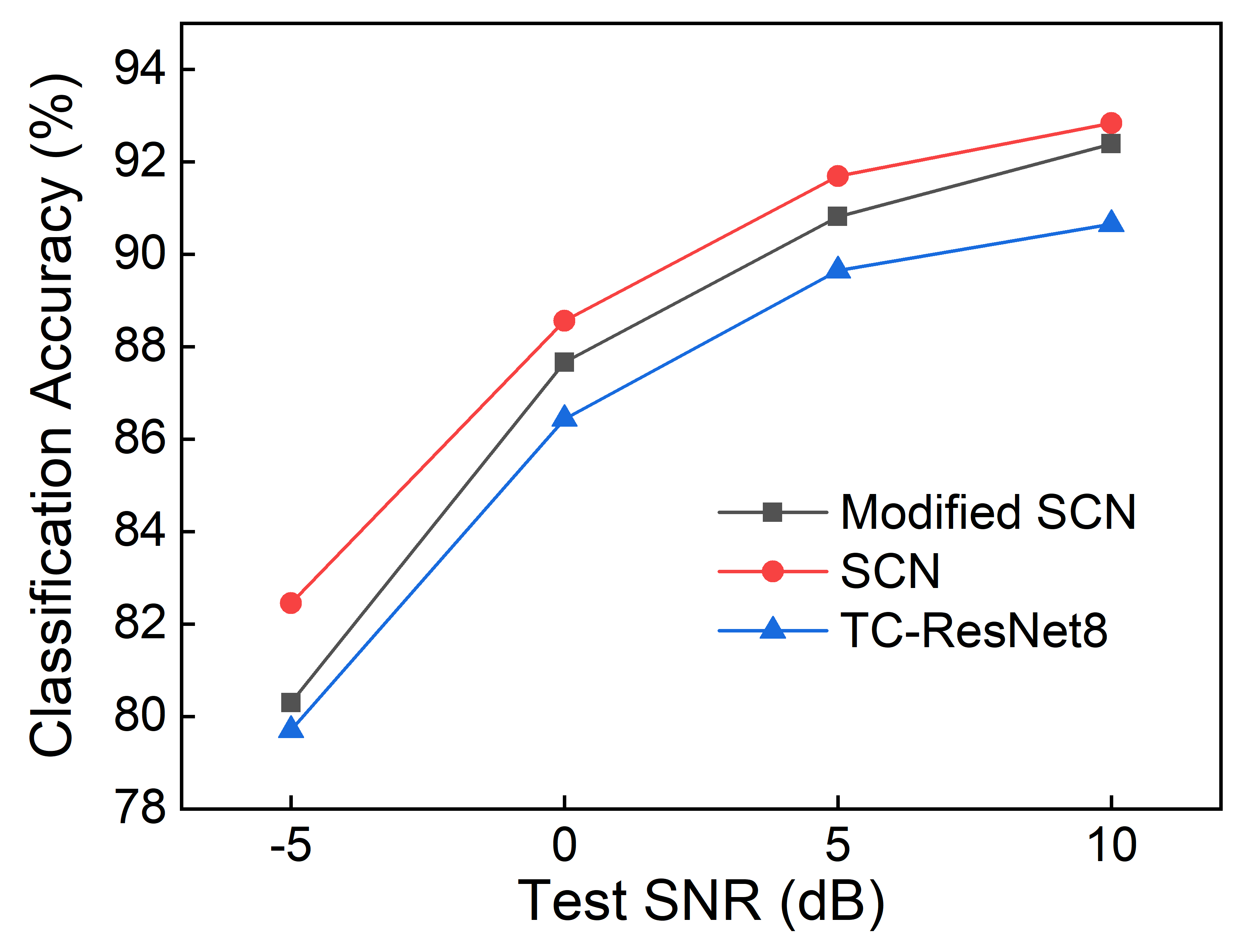}
  \caption{Performance of networks under known test noise conditions. The network accuracy steadily decreases as the amount of noise in the samples is increased.}
  \label{fig:acc_known_noise}
\end{figure}

In this case, the noise files in miscellaneous category are available during training, validation and test phase i.e. we can train the model to learn the nature of the noise distribution used and give close to state-of-the-art performance. Random chunks from the noise files are added to the keyword signals at a SNR value chosen randomly between [-5dB, +10dB]. The classification accuracies of the different networks are plotted against the noise spectrum in Fig. \ref{fig:acc_known_noise}.

As observed from the results of Fig. \ref{fig:acc_known_noise}, even the state-of-the-art KWS networks are susceptible to high noise, evident from the $\sim$ 10\% fall in accuracy as test noise level increases to -5dB. 

\subsection{Network performance when noise conditions are unknown}

In this case, the miscellaneous noise files are only available in the validation and test phases i.e. while training the noise distribution used in the inference stage is unknown. Hence to train the models under noisy conditions, we inject the training dataset with a random mixture of white and pink noise sampled randomly between [-5dB, +10dB].

\begin{figure}[!htbp]
\centering
  \includegraphics[scale=0.3]{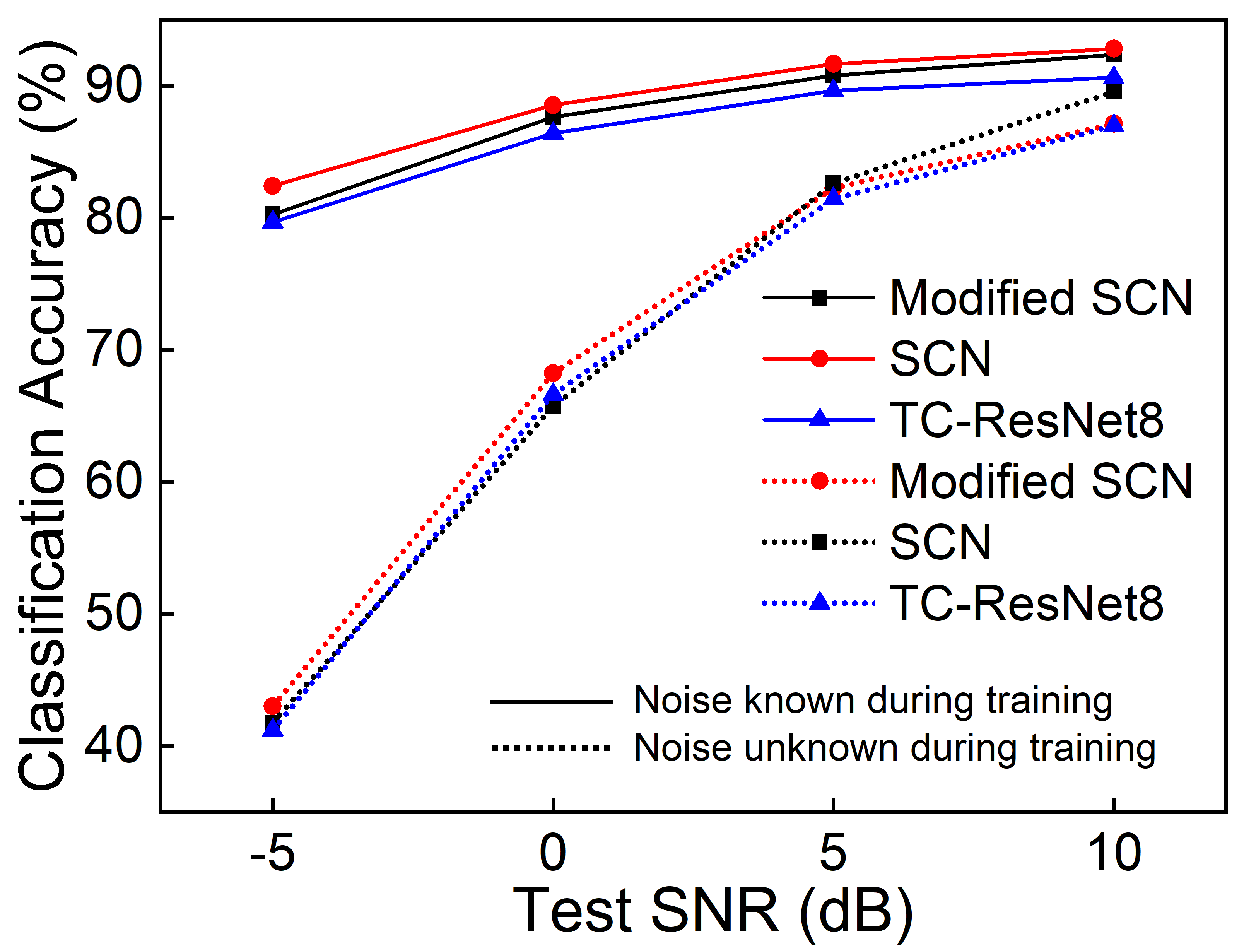}
  \caption{Comparison of performance when noise conditions are known (solid) and unknown (dotted). At 10dB there is a small discrepancy between the performance in the two conditions, but at -5dB there is almost a 40\% difference between the network performance in the two different conditions. }
  \label{fig:acc_unknown_noise}
\end{figure}

\begin{figure}[!htbp]
\centerline{\includegraphics[width=3.4in]{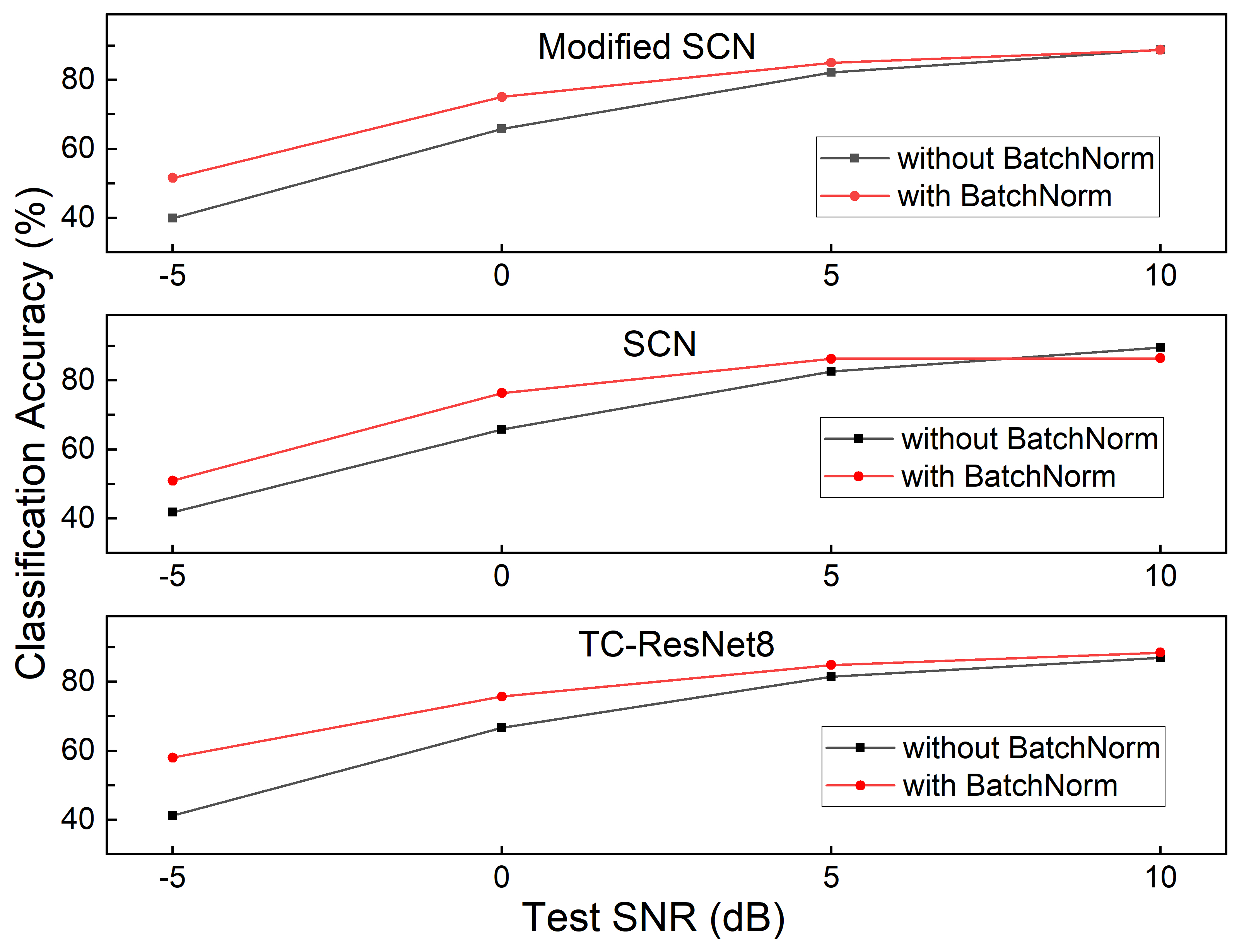}}
\caption{Comparison of accuracy before and after implementing BatchNorm technique. The BatchNorm method is able to significantly boost the performance of all the models under high unknown test noise conditions.}
\label{fig:acc_batchnorm}
\end{figure}

The contrast in the performance of the networks for the two different conditions can be seen in Fig. \ref{fig:acc_unknown_noise}. Though the networks perform satisfactorily under moderate noise ($\sim$10dB range), the performance deteriorates catastrophically under severe noise conditions. To mitigate this, we enable the networks to learn the batch normalized statistics of the validation and test datasets during the corresponding phases rather than depend on the parameters learned during the training phase. The change in the performances of the networks after implementing this is encapsulated in Fig. \ref{fig:acc_batchnorm}.

In Fig. \ref{fig:acc_batchnorm}, though the performance remains almost similar in moderate noise, there is a steady improvement in the final accuracy as we move towards the higher end of the noise spectrum. The SCN models record an improvement of $\sim$10\% and the TC-ResNet model shows a massive rise in accuracy of $\sim$20\% at -5dB test SNR. 

\section{Conclusions and Future Work}

We provide an extensive characterization of the performance of popular KWS networks under heavy noise, and our results show how the existing architectures fail to deliver satisfactory results under non-ideal conditions. We also observe that if the noise in the test phase is not known, training the network by injecting white/pink noise in the training phase performs satisfactorily under moderate noise but fails catastrophically under severe noise conditions. To create networks that perform better under such situations, new models may need to be created.

One solution might be to increase the number of weights and/or layers in the networks and train them on much larger and varied datasets, which also contain an appreciable amount of noise injection. But then this will be contrary to our motive of building networks with small memory footprints. And even though we train the networks using noisy signals, the performance is still sub-par at best. Hence, to create networks that perform better under such situations, new models and algorithms may need to be created. Our BatchNorm algorithm takes one step in that direction by achieving significant enhancement in classification at noisy conditions.


Further improvements to the BatchNorm technique and hardware support for UltraTrail \cite{bernardo2020ultratrail} of the aforementioned techniques and networks is left as future work.

\bibliography{ref.bib}
\bibliographystyle{splncs04}

\end{document}